\documentclass[12pt]{iopart}
\usepackage{iopams}
\usepackage[utf8]{inputenc}
\usepackage{graphicx}
\usepackage[]{subfigure}
\usepackage{enumitem}
\usepackage{epstopdf}
\begin{document}

\title{Nonlocal effects in negative triangularity TCV plasmas}

\author{G. Merlo$^1$, Z.Huang$^2$ ,  C. Marini$^3$, S. Brunner$^4$, S. Coda$^4$, D. Hatch$^5$, D. Jarema$^6$, F. Jenko$^{1,6}$, O. Sauter$^4$ and L. Villard$^4$}
\address{$^1$Oden Institute for Computational Engineering and Sciences, The University of Texas at Austin, Austin, TX 78712, USA}
\address{$^2$ Plasma Science and Fusion Center, Massachusetts Institute of Technology, 77 Massachusetts Avenue, NW17 Cambridge, MA 02139, USA}
\address{$^3$ General Atomics, San Diego, California 92121, USA}
\address{$^4$ Ecole Polytechnique F\'ed\'erale de Lausanne (EPFL), Swiss Plasma Center (SPC), CH-1015 Lausanne, Switzerland}
\address{$^5$ Institute for Fusion Studies, The University of Texas at Austin, Austin, TX 78712, USA}
\address{$^6$ Max-Planck-Institut f\"ur Plasmaphysik, Boltzmannstr. 2, D-85748 Garching, Germany}
\ead{gmerlo@ices.utexas.edu}

\date{\today}

\begin{abstract}
Global gradient driven GENE gyrokinetic simulations are used to investigate TCV plasmas with negative triangularity. Considering a limited L-mode plasma, corresponding to an experimental triangularity scan, numerical results are able to reproduce the actual transport level over a major fraction of the plasma minor radius for a plasma with $\delta_{\rm LCFS}=-0.3$ and its equivalent with standard positive triangularity $\delta$.  For the same heat flux, a larger electron temperature gradient is sustained by $\delta<0$, in turn resulting in an improved electron energy confinement. Consistently with the experiments, a reduction of the electron density  fluctuations is also seen. Local flux-tube simulations are used to gauge the magnitude of nonlocal effects. Surprisingly, very little differences are found between local and global approaches for $\delta>0$, while local results yield a strong overestimation of the heat fluxes when $\delta<0$. Despite the high sensitivity of the turbulence level with respect to the input parameters, global effects appear to play a crucial role in the negative triangularity plasma and must be retained to reconcile simulations and experiments. Finally, a general stabilizing effect of negative triangularity, reducing fluxes and fluctuations by a factor dependent on the actual profiles,  is recovered.
\end{abstract}

\maketitle

\section{Introduction}
A known path towards controlling microturbulence and, as a result, improving the energy confinement in tokamaks is via plasma shaping. Modifying the magnetic equilibrium allows to influence the behaviour of the plasma down to the microscopic scale, in turn potentially reducing heat and particle transport levels.

Along these lines, of particular interest is the intriguing observation, originally made in the TCV tokamak, that flipping the sign of the triangularity of the Last Closed Flux Surface (LCFS) $\delta_{\rm LCFS}$ from positive to negative leads to a significant improvement of the electron energy confinement. The first experimental observation of this kind was reported in Pochelon {\it et al.}\cite{Pochelon_1999}, and a detailed comparison was discussed by Camenen {\it et al.} in Ref.~\cite{Camenen_NF}. For the experimental conditions of \cite{Camenen_NF}, an L-mode limited plasma, the same electron temperature and density profiles were obtained injecting half the heating power when $\delta_{\rm LCFS}$ was inverted from +0.4 to -0.4. Thanks to power balance analysis, this result was interpreted as a better electron energy confinement, by a factor of two, at all radial locations in the outer half of the confined plasma. With the dominant microinstability being Trapped Electron Modes (TEMs), it was suggested that the influence of $\delta$ on the toroidal precessional drift could potentially explain the observed turbulence reduction~\cite{Marinoni_PPCF}. Nonetheless, given the penetration depth of $\delta$, which is finite and large only at the LCFS and then rapidly vanishes moving towards the magnetic axis, the origin of the observed radially uniform improvement remained unexplained. 

Research on negative triangularity has been reinvigorated in recent years thanks to new experimental data confirming its advantages. New and more detailed TCV measurements confirm that $\delta<0$ induces a reduction of transport together with both electron temperature and density fluctuations \cite{Fontana_NF, Huang_coda}. Similar and complementary experimental investigations have been carried out in the DIII-D tokamak, where discharges with negative triangularity showed H-mode-like confinement and high normalized $\beta$ with L-mode-like edge pressure profiles and no ELMs \cite{Marinoni_PoP,austin_PRL}. These plasmas show, in the absence of a transport barrier, the same global performance as a positive triangularity ELMy H-mode discharge with the same plasma current, elongation and area. A reduction of fluctuations and diffusivities has been observed as well, confirming that negative triangularity is a viable candidate for reactor scenarios with high confinement \cite{kikuchi}.  This is particularly appealing since such a reactor scenario would avoid the well-known Edge Localized Modes (ELM) problems associated with H-mode operation.  

Various efforts have been made with the aim of understanding the origin of this confinement increase. A first attempt at reproducing this experimental observation was carried out in Ref.~\cite{Marinoni_PPCF}, where local gyrokinetic simulations were  performed at different radial locations considering the two original TCV experimental conditions. In agreement with a beneficial effect of negative $\delta$ on TEM turbulence, the simulated heat flux through the electron channel was found smaller when $\delta<0$, but only for runs performed  assuming conditions of the outermost radial locations, i.e. where triangularity is large in magnitude, and without being able to reproduce the turbulence suppression observed in the core. Further local simulations have been discussed in Ref.~\cite{Merlo_PPCF}, where they were used to investigate the hypothesis of a triangularity-dependent profile stiffness \cite{Sauter_stiff}. Even in this latter case, local runs were unable to reproduce the experimental transport level, which was largely overestimated especially when assuming core conditions. The common conclusion of Refs.~\cite{Marinoni_PPCF} and \cite{Merlo_PPCF} was that  negative $\delta$ indeed exerts a beneficial effect on TEMs, however the observed radially uniform improvement cannot be reproduced with local simulations and finite machine size effects need to be included in order to reproduce the measured heat flux. In Ref.~\cite{Merlo_PPCF} it was also noted that impurities and collisional effects in particular played a significant role and therefore needed to be retained to ensure sufficiently realistic numerical simulations. Finally, in Ref.~\cite{Merlo_POP} the effect of additional ion heating was investigated considering TCV plasmas with $T_e\simeq T_i$, aiming at understanding if the effect of negative triangularity is limited to TEM dominated regimes or not. A beneficial effect also in ITG dominated plasmas was observed.

All the aforementioned works featured local analyses, and to our knowledge the only global gyrokinetic simulation results  available in literature are discussed in Ref.~\cite{Camenen_NF}, where however only linear stability is discussed. This motivated the present work, where we report the first nonlinear global gyrokinetic simulations investigating the effect of negative triangularity. To this end we have use the GENE code \cite{Jenko_GENE,Goerler_GENE}. This paper focuses on TCV parameters, for which one expects global effects to be particularly relevant owing to its large normalized Larmor radius $\rho^*$ \cite{McMillan}. We nonetheless point out that it is {\it a priori} not obvious to predict how important such effects are. See for instance Refs.~\cite{Mariani_2016,Merlo_2018} which consider other TCV conditions (standard positive triangularity shapes) in which global effects appear to be less relevant, or Ref.~\cite{banon} showing a unexpected global effects for machines of bigger size. 

It turns out that, for the specific L mode TCV limited discharges considered in this paper, global effects can play a crucial role and their significance depend on the plasma shape. For the $\delta>0$ case, minor differences are found between local and global results, which are compatible with profile stiffness. Both approaches agree well with the experimental power balance over a major fraction of the plasma minor radius. For $\delta<0$ case however, even accounting for profiles being very stiff, local results overestimate the actual heat fluxes and only when global effects are included is it possible to recover the correct transport level. Consistently with experiments, a reduction of fluxes and fluctuations is recovered in global simulations with $\delta<0$ compared to $\delta>0$. The exact details of this reduction are dependent on the specific plasma profiles.

The remainder of this paper is organized as follows. Section \ref{sec:exp} presents at a high level the experimental discharge that we considered as a basis for our work. Section \ref{sec:gene} briefly summarizes the numerical tools that we have used. Sections \ref{sec:linear} and \ref{sec:nonlinear} discuss respectively linear and nonlinear results obtained for the experimental conditions. A comparison of local vs. global simulation results is carried out in Section \ref{sec:global_effects}, while a numerical experiment to identify the impact of profiles is discussed in  Section \ref{sec:2spec}. Discussion and conclusions are drawn in Section \ref{sec:concl}. 

\section{TCV experimental discharge}\label{sec:exp}
\begin{figure}
    \centering
    \includegraphics[width=5.3cm]{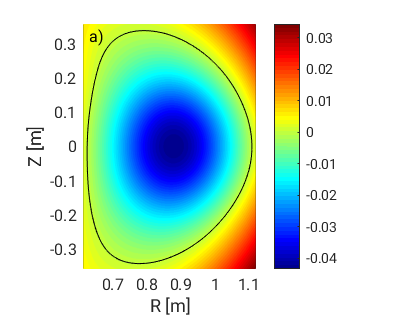}
    \includegraphics[width=5cm]{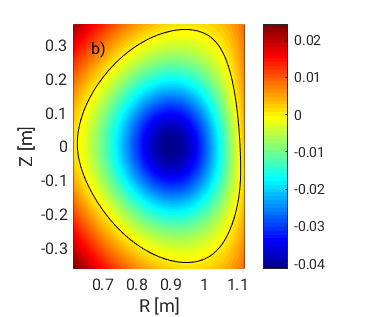}\\
    \includegraphics[width=5cm]{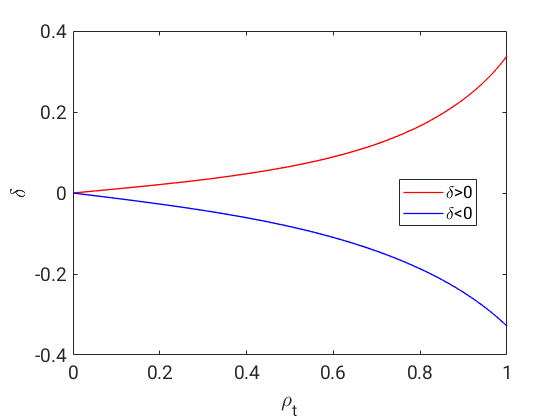}
    \includegraphics[width=5cm]{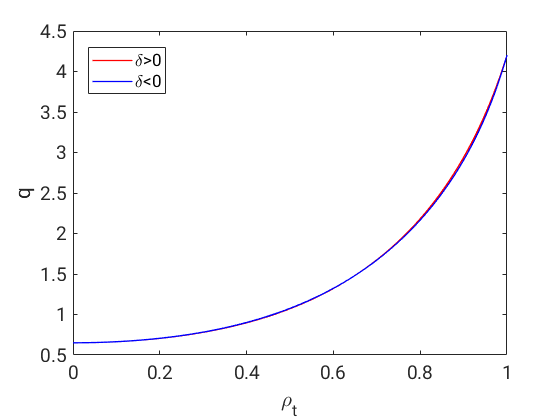}
    \caption{Magnetic geometry of TCV discharge $\#49451$. Shown are the value of the normalized magnetic flux for $(a)$ $\delta >0$ and $(b)$ $\delta<0$, $(c)$ triangularity and $(d)$ safety factor profiles.}
  \label{fig:MHD_eq}
\end{figure}

\begin{figure}
    \centering
    \includegraphics[width=5cm]{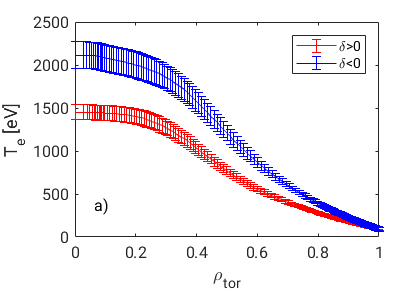}
    \includegraphics[width=5cm]{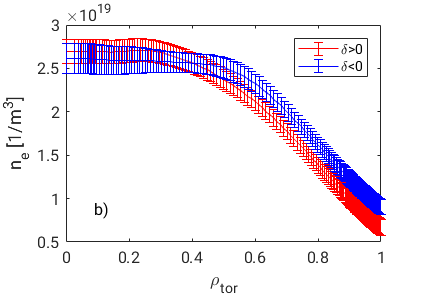}\\
    \includegraphics[width=5cm]{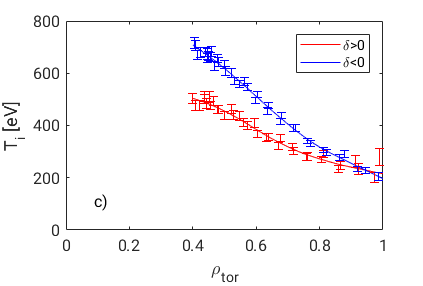}
    \includegraphics[width=5cm]{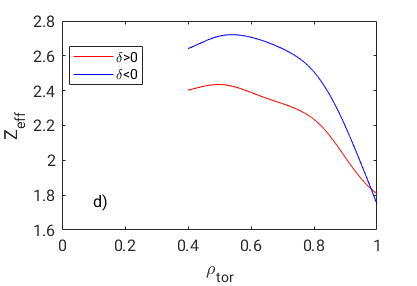}
    \caption{Plasma profiles of $(a)$ electron temperature $T_e$, $(b)$ electron density $n_e$,  $(c)$ ion temperature and $(d)$ $Z_{\rm eff}$ for positive (red) and negative (blue) triangularity.}
  \label{fig:profs}
\end{figure}

In order to investigate the effect of triangularity on turbulent transport and fluctuations, we consider TCV discharge $\#49451$. This discharge is a limited L-mode plasma in which the triangularity of the Last Closed Flux Surfaces (LCFS) is continuously varied from negative, $\delta_{\rm LCFS}=-0.3$, to positive, $\delta_{\rm LCFS}=0.3$. A careful experimental effort was made to keep all plasma parameters, except $\delta$, as unchanged as possible during the plasma evolution. This discharge is however carried out at constant heating power power, in contrast to the original ones described in \cite{Camenen_NF} and used as a basis for gyrokinetic analysis in \cite{Marinoni_PPCF, Merlo_PPCF}. This results in clear differences in the  plasma temperature and density profiles, such that it is not justified to assume that profiles are the same between the two shapes, as was done in the aforementioned analyses. We therefore choose for our analysis the two instants t=0.6s and t=1.5s, as they are the ones where overall we find the smallest differences in the geometrical coefficients besides flipping the sign of the edge triangularity. The experimental magnetic geometries, together with the radial profiles of safety factor $q$ and triangularity are shown in Figure \ref{fig:MHD_eq}. Here $\rho_{tor}$ is the radial coordinate based on the toroidal flux $\Phi$, $\rho_{tor}=\sqrt{\Phi/\Phi_{\rm LCFS}}$. The magnetic equilibria have been reconstructed using CHEASE \cite{CHEASE}.
Plasma profiles, the experimental measurements together with the associated fits, are shown in Figure \ref{fig:profs}.
Nearly identical density profiles are obtained for the two shapes, as shown in Figure \ref{fig:profs}(b), although the negative triangularity has slightly higher density outside $\rho_{tor}$=0.5, resulting in a somewhat lower logarithmic scale length $a/L_{n_e}$. A significantly higher central electron temperature, by about 30$\%$ to 50$\%$, is observed when  $\delta<0$, as a result of reduced turbulence and associated transport level. The corresponding electron temperature gradient is larger at all radial locations as well. A similar behavior is also found for the ion temperature profile. One notes as well that an approximately 15$\%$ higher effective charge $Z_{\rm eff}$ is observed for $\delta<0$. For a more detailed description of the experimental discharges and measurements, the interested reader is referred to Ref.~\cite{Huang_coda}.

\begin{figure}
    \centering
    \includegraphics[width=6cm]{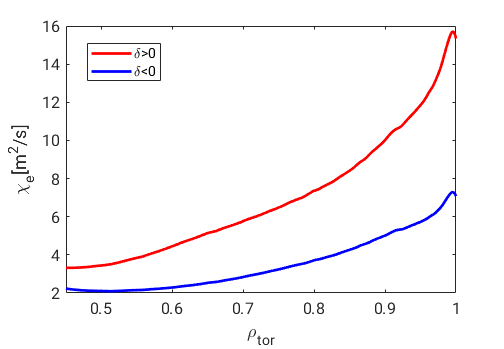}
    \caption{Profiles of electron heat diffusivity obtained from experimental power balance.}
  \label{fig:chie}
\end{figure}
Finally, Figure \ref{fig:chie} shows the electron heat diffusivity, experimentally evaluated by attributing to the electron channel the Ohmic and external (ECRH) heating. The equipartition flux is carried by the main ions. A clearly improved confinement, reflecting the higher electron temperature gradient for $\delta<0$, can be seen, with a reduced diffusivity at all radii by approximately a factor of 2 (1.7 at $\rho_{tor}=0.5$ and 2.2 at $\rho_{tor}=0.95$).

\section{Simulation tool and numerical setup}\label{sec:gene}
All simulations described here have been performed using the gyrokinetic code GENE, using both its local (flux-tube) and global versions. A detailed description of the code can be found e.g. in Refs.~\cite{Jenko_GENE, Goerler_GENE}. Here we simply recall the most relevant code features. GENE adopts a field aligned coordinate system $(x,y,z)$ to discretize the configuration space, while $(v_{\parallel},\mu)$ are used as velocity variables. Here $x$ stands for the radial, $y$ for the binormal and $z$ is the straight-field line poloidal angle, parametrizing the position along a given field line. The variable $\mu=mv_{\perp}^2/2B$ represents the magnetic moment, while $v_{\parallel}$ and $v_{\perp}$ are the components of velocity respectively parallel and perpendicular to the magnetic field; $m$ is the mass of the particle and $B$ the local magnitude of the magnetic equilibrium field $\mathbf{B}$. A block-structured discretization of the velocity space \cite{jarema_16} has been used for global runs, matching in each block the resolution used for the corresponding local simulations.

The considered plasmas are deuterium discharges, with carbon ($Z_C=6$) the dominant impurity. Three fully kinetic species have thus been considered in the GENE runs: deuterium ions, electrons and carbon ions. Deuterium and carbon concentrations are estimated so as to respect quasineutrality, i.e.~$n_D +Z_C n_C=n_e$ and in agreement with the experimentally measured value of $Z_{\rm eff}=\sum_i n_i Z_i^2 / n_e$, where the sums are over all ion species. The experimental value of $\beta_e=2\mu_0p_e/B_0^2$ is used to include electromagnetic effects.  Inter- and intra-species collisions are retained and all collision frequencies are consistently derived from the value of $\nu_{ei}$ and local values of temperature and density. In particular, $\nu_{ei}=\sum_i{3\sqrt{\pi}/4\tau_{e,i}}$, where $\tau_{e,i}$ is the electron collision time with the $i$-th ion species,  $\tau_{e,i}=3(2\pi)^{3/2}\epsilon_0^2T_e^{3/2}m_e^{1/2}/n_iZ_i^2e^2 \log\Lambda$, $m_e$ is the electron mass and $\log\Lambda$ is the Coulomb logarithm. 

Global simulations also require one to use source terms in order to prevent relaxation of the background profiles. The simulations described here are gradient-driven and make use of a Krook-type operator to prevent the relaxation of the background temperature and density profiles. The associated relaxation rate $\gamma_{K}$ has been set to approximately 1/10 of the linear growth rate as in \cite{lapillonne} and results are found not to be significantly affected when $\gamma_{K}$ is varied. More realistic and computationally challenging flux-driven runs are left for future work. 

\section{Linear simulation results for experimental conditions}\label{sec:linear}
Global simulations have been performed covering the radial annulus $0.4<\rho_{tor}<0.95$. This choice of the radial domain is dictated by the available ion temperature and density profiles measurements. Charge eXchange Recombination Spectroscopy, routinely used in TCV to measure Carbon and thus infer main deuterium profiles, is available only over the said radial extent. No information is available for positions further inside in the plasma. At the same time, a large error bar affects measurements close to the LCFS, thus this region is also excluded from the computational domain. 

\begin{figure}[h]
    \centering
    \includegraphics[width=8cm]{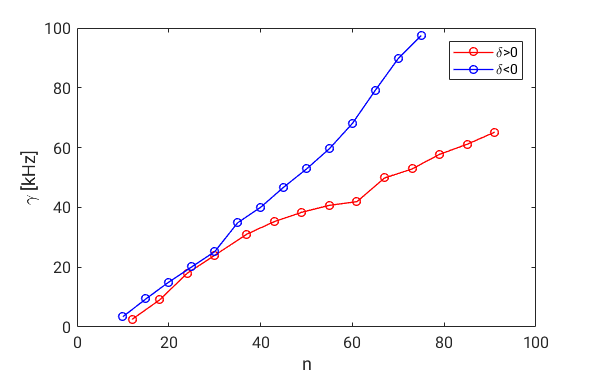}
    \caption{Growth rates of the most unstable mode as a function of the toroidal mode number $n$ evaluated with global simulations, in blue for negative triangularity and in red for positive.}
  \label{fig:linear_global}
\end{figure}
Figure \ref{fig:linear_global} compares growth rates between positive and negative triangularity. We plot results as a function of the toroidal mode number $n$ (values correspond to the ion scale, $k_y\rho_s<2$ as these are the scales we will consider in nonlinear runs) and in SI units such as to facilitate the comparison. Based on their negative frequency in GENE normalizations, all modes are identified as Trapped Electron Modes (TEM). One notes that for very small wave numbers, $n\simeq20$ or $k_y\rho_s\simeq0.3$ where most of the nonlinear transport happens, growth-rates are similar, if not slightly larger for $\delta <0$, while for higher $n$ negative triangularity is in fact significantly more unstable. While this might appear in contradiction with the stabilizing role of negative triangularly seen in the experiments, a comparison between the growth rates obtained for the two shapes is non trivial. Modes peak at different radii reflecting the different driving gradients, and therefore any conclusion based on linear global runs should be taken with care. For instance, the similar growth rates seen at low $n$, where for both shapes the modes  peak around $\rho_{tor}$=0.65 appear consistent with the fact that the discharges have the same heating power (and larger gradients characterize the $\delta<0$ case).  Higher $n$ modes instead peak more outwards when $\delta<0$, where for a given shape the background gradients are larger, but they are in fact not contributing significantly to the overall transport level which is mostly due to low $n$ modes.

\begin{figure}
    \centering
    \includegraphics[width=6cm]{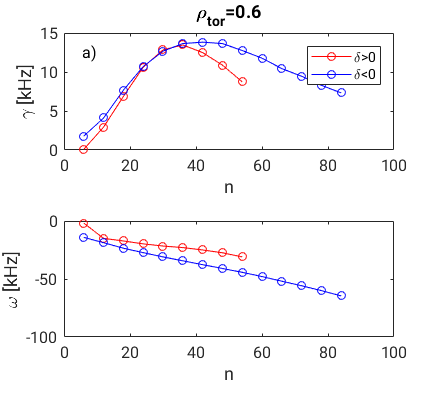}
    \includegraphics[width=6cm]{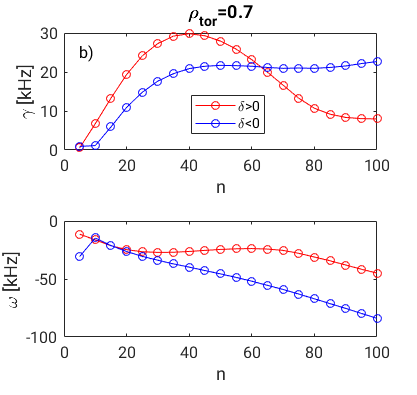}
    \includegraphics[width=6cm]{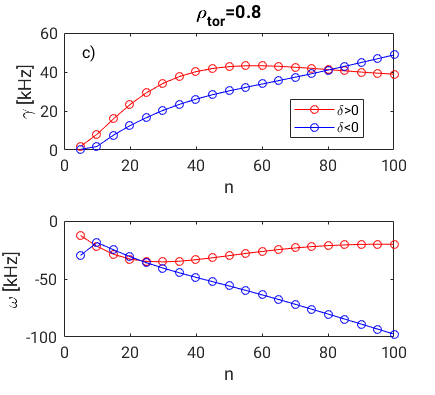}
    \includegraphics[width=6cm]{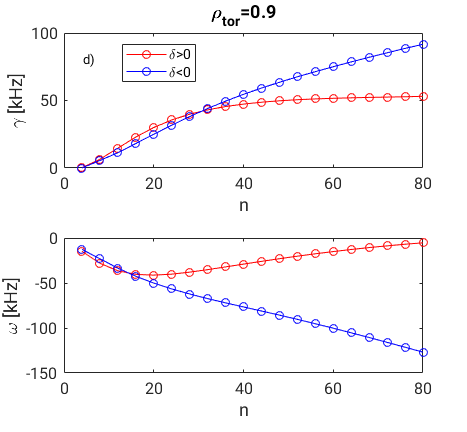}
    \caption{Growth rates (top panels) and frequency (bottom panels) evaluated with local simulations at different radial locations, in blue for negative triangularity and in red for positive.}
  \label{fig:linear_local}
\end{figure}
To help in understanding the results of global simulations, we have also performed linear scans at different radii,  presented in  Figure \ref{fig:linear_local}. We limited this analysis to $\rho_{tor}$=0.6, 0.7, 0.8 and 0.9. We did not consider more radially inward positions because for $\rho_{tor}<0.5$ the nominal profiles are found to be stable. Again at all radii, TEMs are the dominant instability. For low $n$, $ \delta<0$ is characterized by similar or smaller growth rates, while again we find that high $n$ modes are more unstable when triangularity is negative. 

\section{Nonlinear simulation results for experimental conditions}\label{sec:nonlinear}
\subsection{Transport level}
\begin{figure}[h]
    \centering
    \includegraphics[width=6cm]{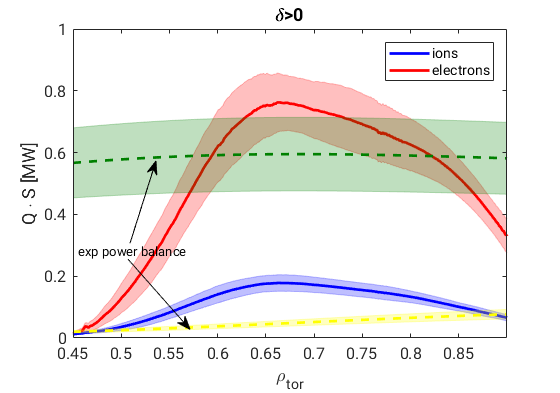}
    \includegraphics[width=6cm]{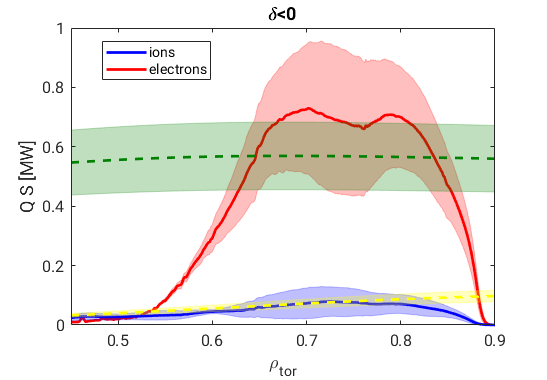}
    \caption{Simulated electron (red) and ion (blue) heat flow for $(a)$ positive and $(b)$ negative triangularity. Experimental power balance estimates are shown with dashed lines, in green for electrons and in yellow for ions. Shaded areas indicate error bars.}
  \label{fig:global_transport}
\end{figure}

The transport obtained from nonlinear global GENE runs is shown in Figure \ref{fig:global_transport}, where numerical results are compared to the experimental power balance. As already mentioned, the experimental power balance is estimated attributing to the electron channel the entire heating power (Ohmic + ECH) and to the ions only the remaining  equipartition contribution. As such we do not have a rigorous estimation of the associated error bar; a 20$\%$ uncertainty has been chosen as an approximate value. Solid red (resp. blue) lines indicate the electron (resp. ion) heat flow, while green (resp. yellow) the corresponding experimental power balance estimates. The simulated flux accounts for both the electrostatic and electromagnetic contribution, the latter being as expected smaller by at least one order of magnitude. Carbon heat flux, also negligible, is not shown for simplicity. In order to evaluate error bars from simulations, time traces are divided in disjoint intervals of approximate length 80 $a/c_s$ and averages are computed in each interval. The error bars is then takes as one standard deviation of the mean values.
We observe a good agreement between simulations and experiments over the region $0.6\leq\rho_{tor}\leq0.9$, where the simulated electron heat flow matches within error bars the measured one for both shapes. Over the same region, the overall flux, sum of ion and electron channels, in fact slightly overestimates for the $\delta>0$ case. For both shapes, simulations don't recover the experimental flux in the innermost part of the radial domain, where GENE results show a rapidly vanishing heat flow in contrast to the finite transport level inferred from the experiment. We have considered a different (larger) radial domain, as well as an increased binormal resolution such as to include possible contributions from the electron scales  but the simulation results were not affected. The same very low transport level is also observed in local simulations, therefore we can exclude it being a numerical artifact caused by the radial boundary condition or our numerical set-up and we can only attribute this simulations-experiments difference to profile stiffness and errors in the reconstruction of the plasma profiles. Indeed the region $\rho_{tor}\leq0.6$ is very close to the inversion radius, which together with the sawtooth activity, makes the profile reconstruction particularly challenging. Furthermore, the transport seen in the experiment is the result of the sawtooth activity, which is not captured by the our gyrokinetic model. We have also carried out simulations without carbon impurities, see also Sec.~\ref{sec:2spec}, and a similar behavior is observed. The inclusion of impurities is nonetheless necessary in order to recover the actual transport level, in particular for the ion channel which would otherwise be significantly overestimated.

\subsection {Fluctuations}
One of the most evident observations made in negative triangularity discharges is the reduction of fluctuation amplitudes, of both electron density and temperature, down to the core of the plasma where triangularity is essentially vanishing. In this section we look at GENE results and at the impact of negative triangularity on fluctuation fields. To this end, we evaluate their root mean square value at the outboard midplane as follows:
\begin{equation}
A(\rho_{tor})=\sqrt{\langle|A(\rho_{tor},\theta=0)|^2}\rangle_t
\label{eq:fluct}
\end{equation}
where $A$ stands for electron temperature or density fluctuation and $\langle \cdot \rangle_t$ indicates a time average. This choice aims at carrying out a qualitative comparison with experimental fluctuation measurements and their radial profiles. A one-to-one comparison with actual measurements requires to make use of specific synthetic diagnostics, similar to what done in e.g.~\cite{rost,Hausammann_2017}. Such tools are already available with the GENE code, \cite{Merlo_thesis}, and are based on the integration of the fluctuating fields over a specific volume such as to reproduce the same instrumental transfer functions as in the experiments. Unfortunately, for the specific discharge we considered, both density (with Phase Contrast Imaging (PCI)) and temperature fluctuation measurements (with correlation Electron Cyclotron Emission) are taken in the core region ($\rho_{tor}=0.5$ or below), where our simulations show a weak turbulence and fail to reproduce the transport level. Therefore, we leave a detailed validation study for future work. We nonetheless point out that, consistently with experiments, we observe lower transport and fluctuation levels for $\delta <0$ at $\rho_{tor} < 0.5$.

\begin{figure}[h]
    \centering
    \includegraphics[width=7cm]{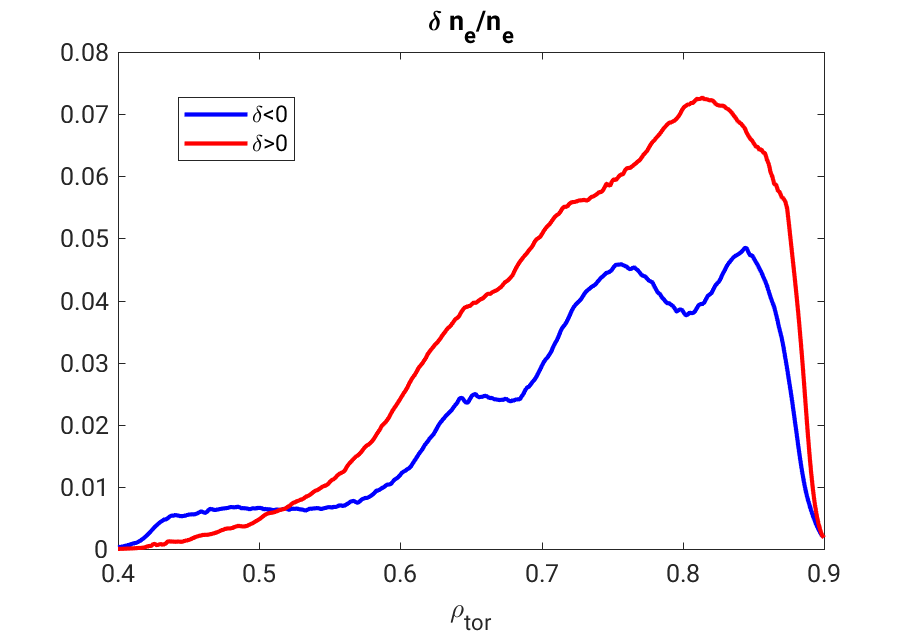}
    \includegraphics[width=6cm]{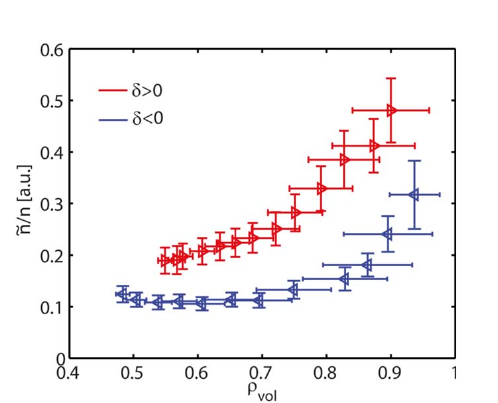}
    \caption{Radial profiles of electron density fluctuations, for positive and negative triangularity. Left plot shows GENE simulation results computed according to \eref{eq:fluct} while right plot shows PCI measurement for a similar shot. Reproduced from Ref.~\cite{Huang_coda}.}
  \label{fig:dens_3s}
\end{figure}

\begin{figure}[h]
    \centering
    \includegraphics[width=7cm]{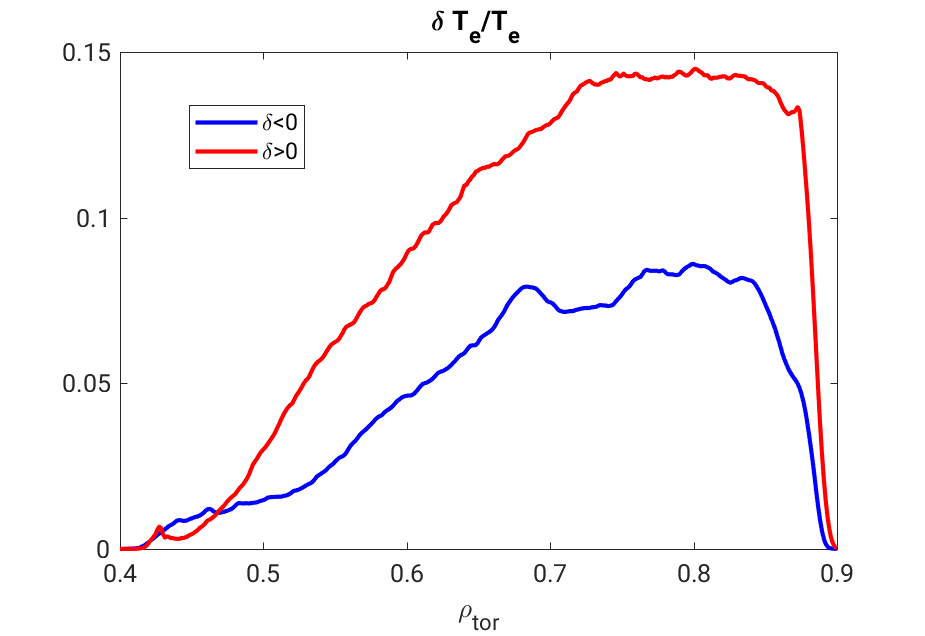}
    \includegraphics[width=7cm]{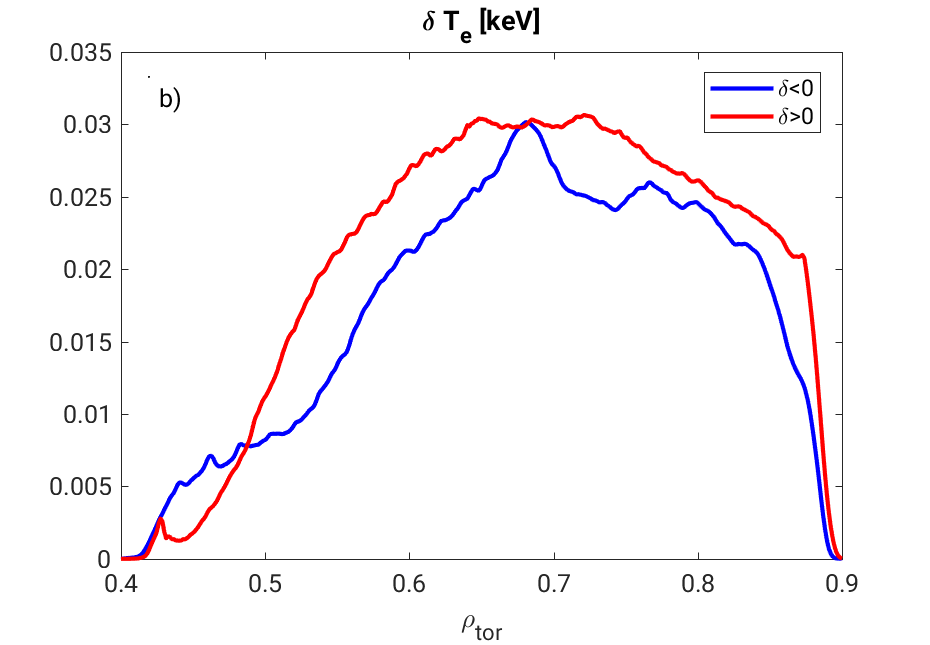}
    \caption{Radial profiles of $(a)$ relative and $(b)$ absolute electron temperature fluctuations, for positive and negative triangularity, computed from GENE simulations using Eq.~\eref{eq:fluct}.}
  \label{fig:temp_3s}
\end{figure}

Figure \ref{fig:dens_3s} shows the behavior of electron density fluctuations, compared to PCI measurements obtained from a similar triangularity scan at similar heating power (TCV $\#49052$, 0.45MW rather than 0.6 MW) \cite{Huang_coda}.
One observes a striking similarity between simulations and experimental data, with a reduction of fluctuations for $\delta<0$ that extends throughout the entire simulated domain. A similar observation can be make looking at the profiles of relative  temperature fluctuation, Figure \ref{fig:temp_3s}(a). 
In this latter case one sees almost a constant reduction by a factor of two at all radii when the shape is changed. It should however be pointed out that most of this difference is caused by the different background temperatures, i.e. higher $T_e$ for $\delta<0$. A consistent difference in the absolute value of the temperature fluctuation nevertheless remains when the triangularity is flipped, as can be seen in \ref{fig:temp_3s}(b), reflecting the lower electron transport level of Figure \ref{fig:global_transport}.

\section{Nonlocal effects}\label{sec:global_effects}
Global effects are generally expected to be relevant for machines of the size of TCV \cite{McMillan}. At the center of our simulation domain, $\rho^*=\rho_s/a$ is approximately $1/140$ (resp. $\sim1/110$) for the positive (resp. negative) triangularity plasma considered in this paper, and therefore are often invoked to explain the frequently observed differences between local (flux-tube) results and experiments. More specifically, in the case of negative triangularity they have been speculated to be a crucial ingredient leading to the observed radially global confinement improvement and, although the more recent DIII-D  observations clearly point out that the effect of $\delta<0$ is not limited to small machines, global effects cannot be ruled out without actually carrying out simulations and measuring their impact. The results presented in the previous section provide almost the ideal scenario for studying finite machine size effects because global simulations are able to reproduce several experimental observations and therefore it is natural to confront them with the corresponding local runs. This section describes such a comparison.

Before discussing the numerical results, we should also point out that accurately reproducing experiments with gradient driven simulation results (both local and global) is normally a challenging effort because of the so-called ``profile stiffness'', i.e. the simulated transport level is extremely sensitive to variations of the input temperature and density gradients. At the same time, measured plasma temperature and density profiles usually present large error bars, such that, for a lot of cases, it turns out to be sufficient to vary the background gradients within experimental error bars in order to recover the observed transport level. This high sensitivity needs to be accounted for also when looking for global effects, since appropriate source terms need to be used to prevent background profiles to relax in global runs, but a minor relaxation is nevertheless to be expected.

\begin{figure}
    \centering
    \includegraphics[width=7cm]{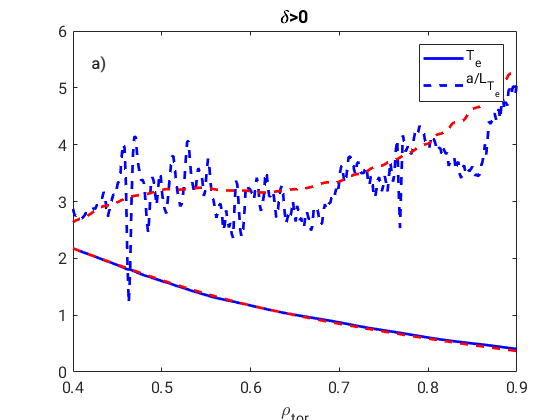}
    \includegraphics[width=7cm]{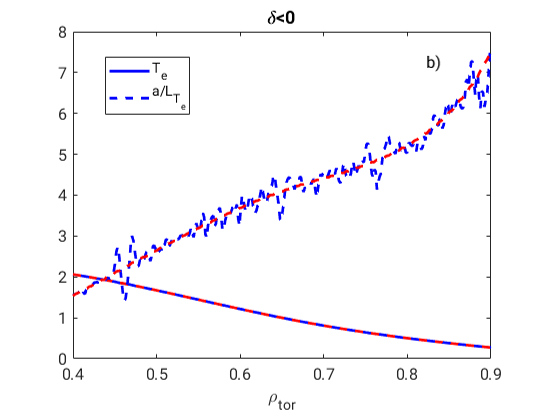}
    \caption{Electron profiles (solid lines) and their logarithmic gradient (dashed lines), used in the simulations for $(a)$ positive and $(b)$ negative triangularity. Red lines indicate the initial profiles while blue lines the time-averaged profiles resulting from gradient-driven runs. Temperatures have been normalized to $T_0$, their values at the center of the simulation domain.}
  \label{fig:global_profs}
\end{figure}

Figure \ref{fig:global_profs} shows the actual electron temperature profiles obtained taking a time average over the second half of the corresponding global simulations. Some relaxation, max 10$\%$, can be seen. To assess the importance of global effects, we have performed a series of local runs, at the position $\rho_{tor}=$0.6, 0.7, 0.8 and 0.9 considering the nominal background profiles. These simulations employ the same resolution as corresponding global runs in all directions, except in the radial one, which has been adjusted depending on the specific radial location.

\begin{figure}[h]
    \centering
    \includegraphics[width=7cm]{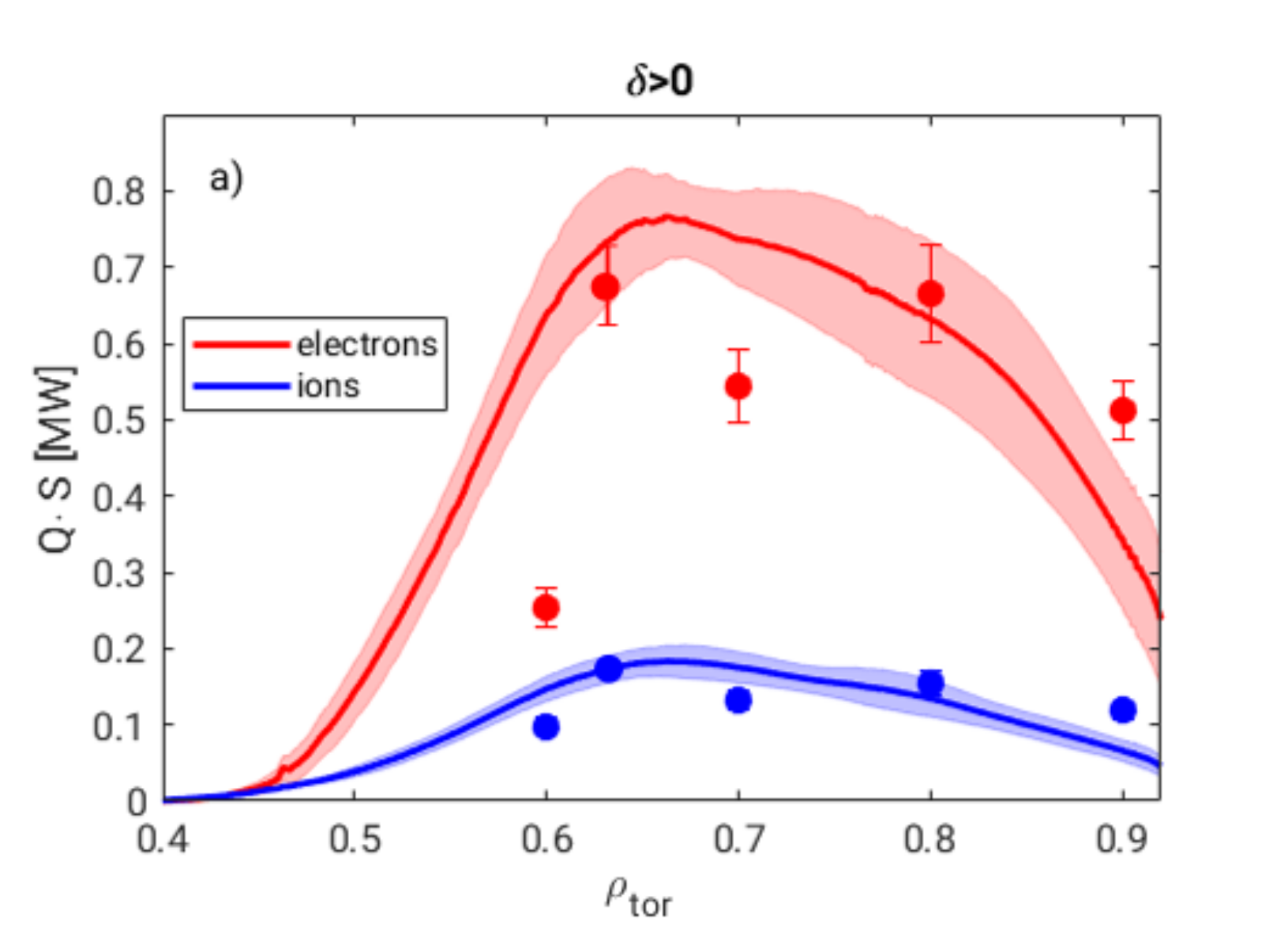}
    \includegraphics[width=7cm]{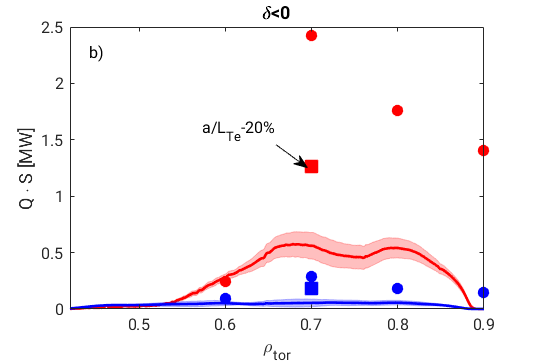}
    \caption{Comparison between transport levels obtained from global (solid curves) and local simulations (full markers),  Global results are the same as in Figure \ref{fig:global_transport}.}
  \label{fig:comparison_local}
\end{figure}

Figure \ref{fig:comparison_local} compares the results of global and local simulations, showing a staggeringly different behavior depending on shape. First, we note how local simulations also predict a very low transport level in the core region ($\rho_{tor}<0.6$) for both shapes, and in this region global effects manifest mostly in the form of turbulence spreading \cite{lin} since global simulations predict a flux that is larger than for the local ones. In the outer region however, ($0.6\leq\rho_{tor}\leq 0.9$), finite machine size effects are very different for the two triangularity cases. Surprisingly, very weak global effects are seen for $\delta>0$, where local simulations are in fact able to essentially recover the experimental power balance of approximately 600 kW. We have specifically considered the position $\rho_{tor}=0.65$, near to the maximum simulated heat flux in the global run for $\delta>0$ and local simulations recover, within error bars, a similar transport level as global results. A large overestimation, by approximately a factor of four, is instead found for all the considered radii when $\delta<0$.
As expected, local results are very sensitive to gradients. Therefore, one has to question whether it is still possible to match the transport level of the negative triangularity plasma with local runs, changing the plasma profiles within  their error bars.  We have investigated this possibility only for the position $\rho_{tor}=0.7$, which is where global effects appear to be largest. 

\begin{figure}[h]
    \centering
    \includegraphics[width=10cm]{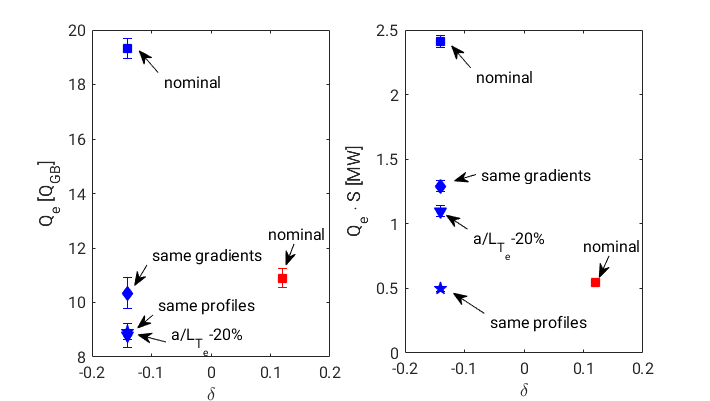}
    \caption{Comparison between transport levels obtained from local simulations centered at $\rho_{tor}=0.7$ of negative (blue) and positive (red) triangularity. Results are shown in GyroBohm units in the left panel and in SI units in the right one.}
  \label{fig:local_scans}
\end{figure}

\begin{table}[h]
    \centering
    \begin{tabular}{c c c c c c c c c c}
    \hline\hline
        & $\kappa$ & $a/L_{T_e}$ & $a/L_{T_i}$ & $a/L_{n_e}$ & $a/L_{n_i}$ & $a/L_{n_c}$ & $T_i/T_e$ & $Z_{\rm eff}$\\
        \hline\hline
    $\delta=-0.14$ &  1.3 & 4.40 & 2.46 & 1.97 & 1.80 & 2.30 & 0.80 & 2.66\\
    $\delta= 0.12$ &  1.25& 3.34 &  3.62 & 2.25 & 2.09 & 2.05 & 0.59 & 1.71\\
    \hline\hline
    \end{tabular}
    \label{table:1}
    \caption{Main plasma parameters characterising the flux-surface $\rho_{tor}=0.7$. In both cases, one has the same value of safety factor $q=1.7$ and magnetic shear $s=1.9$.}
\end{table}
As one can see inspecting Table \ref{table:1}, which lists the essential nominal plasma parameters for the flux-surface $\rho_{tor}=0.7$, the major differences between the two shapes are in the electron temperature gradient and $Z_{\rm eff}$, both higher for the $\delta<0$ case. While the former parameter is expected to cause a higher transport level, the latter has potentially a stabilizing effect because of the increased collisional damping of TEMs. In addition, the higher electron temperature of $\delta<0$ will also introduce different gyroBohm units, as well as a lower collisionality, whose combined effect is not trivial to predict. Thus, in Figure \ref{fig:local_scans} we plot the results of a series of local simulations where we have selectively replaced plasma parameters from their nominal value measured when $\delta<0$ to ones of $\delta>0$. We plot results in both gyroBohm and SI units, aiming at measuring the effects of all parameter differences.
Negative triangularity appears particularly sensitive to variations of the electron temperature gradient, with a reduction by 20$\%$ causing a reduction by more than a factor of two of the electron heat flux, from 2.5 MW down to 1.1 MW (blue triangle in Figure \ref{fig:local_scans}). If all logarithmic gradients and density ratios (i.e. $Z_{\rm eff}$) are matched to the ones of the positive triangularity (blue diamond in Figure \ref{fig:local_scans}) the transport level increases by roughly 20$\%$ with respect to the simple variation of electron temperature scale length, but still strongly reduced with respect to the nominal set-up. A similar transport level between the two shapes (0.49 MW for $\delta<0$ and 0.54 MW for $\delta>0$) is recovered only when the same temperature and density profiles are used for computing all plasma parameters (blue stars). In this latter case not only logarithmic gradients are matched for all species but also the same values of temperature and density, affecting collisionality and electromagnetic effects, are used. This is particularly clear when comparing results in SI to results in gyroBohm units, with the latter showing a transport always lower than the one of $\delta>0$ as soon as the electron temperature gradient is reduced. It should be noted that these local results are in agreement with the previous gyrokinetic results of Ref.~\cite{Merlo_PPCF}, which considered cases where with the same profiles half the heating power (thus half the heat flux) was needed when $\delta<0$. For our plasmas the edge triangularity is smaller that what was considered in Ref.~\cite{Merlo_PPCF}, therefore a smaller $\delta$ is also found at 0.7, and the differences between positive and negative triangularity are much smaller, only 15$\%$, consistent with the local approximation and previous results.

\begin{figure}[h]
    \centering
    \includegraphics[width=6cm]{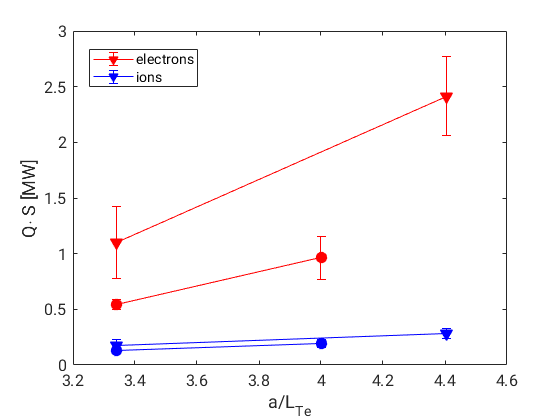}
    \caption{Sensitivity of the electron and ion heat fluxes at the position $\rho_{tor}=0.7$, for positive (squares markers) and negative (triangles) triangularity.}
  \label{fig:local_stiff}
\end{figure}

For the sake of completeness, one has also to consider the possibility that the profiles of positive triangularity are a particularly ``lucky''  fit, such that the agreement between local results and experiments is fortuitous. At the same time, one should also verify that the very weak global effects seen for the $\delta>0$ case are not the result of the specific profiles used. Figure \ref{fig:local_stiff} compares the sensitivity of both positive and negative $\delta$ with respect to variations of the electron temperature gradient, the latter being increased (resp. reduced) by 20$\%$ when $\delta>0$ (resp. $\delta<0$), with the goal of mimicking a systematic error in the fitting procedure.
As expected, the electron heat flux increases with $a/L_{T_e}$ both shapes, but for $\delta>0$ it does not reach similar values as the one obtained for the nominal $\delta<0$ case. Therefore, we can only conclude that global effects appear to be stronger when $\delta<0$. 
Naturally the question of why global effects are more pronounced when $\delta<0$ remains. As already pointed out, a naive measure of global effects is given by $\rho^*$, the ratio of the sound Larmor radius $\rho_s$ to the device minor radius, and this parameters is larger for $\delta<0$ implying stronger global effects. In Ref.~\cite{McMillan}, the ratio between the Larmor radius and a characteristic gradient profile width $\Delta$ was proposed as an alternative measure of global effects. Such a definition is in fact not applicable to realistic plasma profiles, for which it is impossible to define a $\Delta$. Therefore, in agreement with \cite{banon} we use the alternative definition $\rho^*_{\rm eff}=\rho_s/L_{T_e}$, replacing $\Delta$ with the electron temperature gradient scale length. 
\begin{figure}[h]
    \centering
    \includegraphics[width=5.cm]{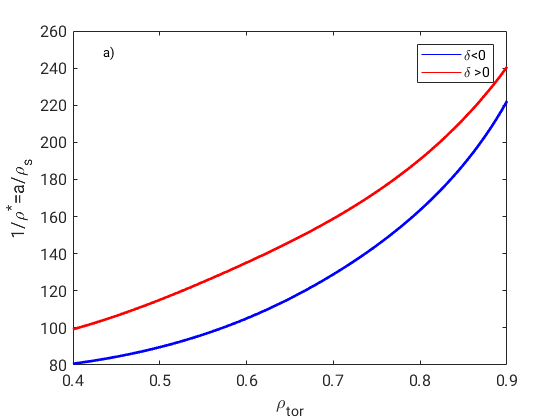}
    \includegraphics[width=5.cm]{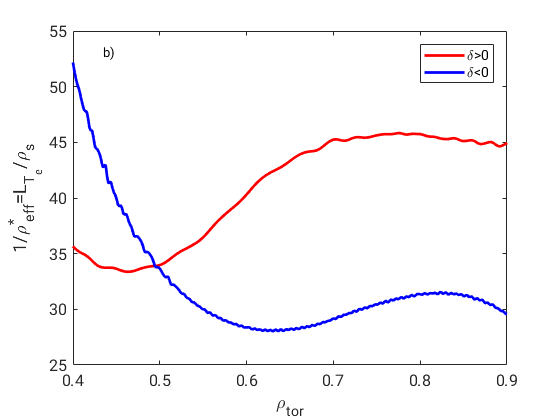}
    \includegraphics[width=5.cm]{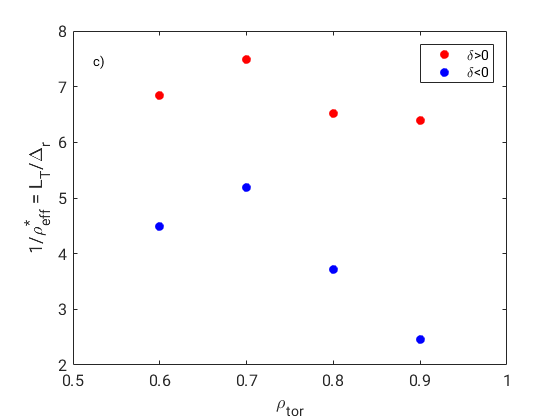}
    \caption{Ratios of $(a)$ minor radius to local Larmor radius ($1/\rho^*$),  $(b)$ electron temperature logarithmic gradient and local Larmor radius ($1\rho^*_{\rm eff}$) and (c) electron temperature logarithmic gradient to radial correlation length ($1\rho^*_{\rm eff}$) for the two TCV conditions considered here.}
  \label{fig:rhostar}
\end{figure}
The inverse of those two quantities are plotted in Figure \ref{fig:rhostar}, where we also show the ratio between the electron temperature gradient scale length and the radial correlation length of turbulence. The latter is evaluated as the correlation of the electrostatic potential at the outboard mid plane. All these metrics show that stronger global effects are to be expected when $\delta<0$, in agreement with our simulations.

\section{A numerical experiment addressing the effect of triangularity}\label{sec:2spec}
\begin{figure}[h]
    \centering
    \includegraphics[width=6cm]{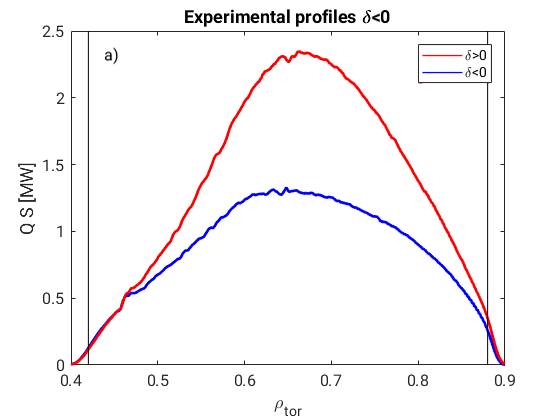}
    \includegraphics[width=6cm]{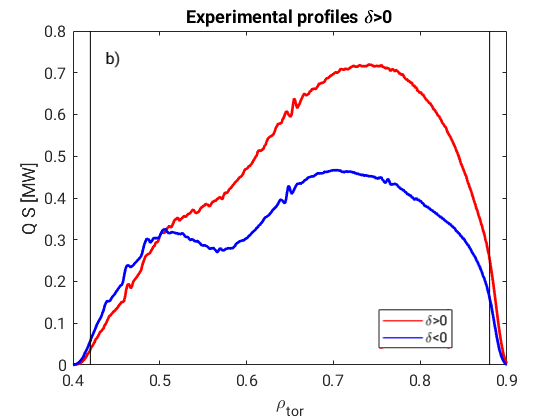}
    \caption{Total heat flux in MW as a function of radius for (red curves) positive and (blue curves) negative triangularity. Experimentally measured profiles of the $\delta <0$ case are used as input for the simulations on the left plot, while profiles from the $\delta >0$ case are used for the right panel.}
  \label{fig:global_Q_red}
\end{figure}

\begin{figure}
    \centering
    \includegraphics[width=6cm]{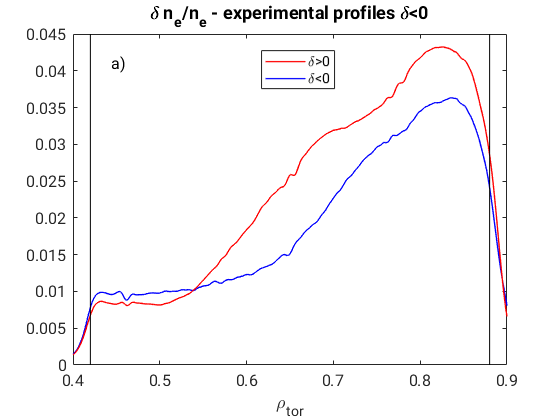}
    \includegraphics[width=6cm]{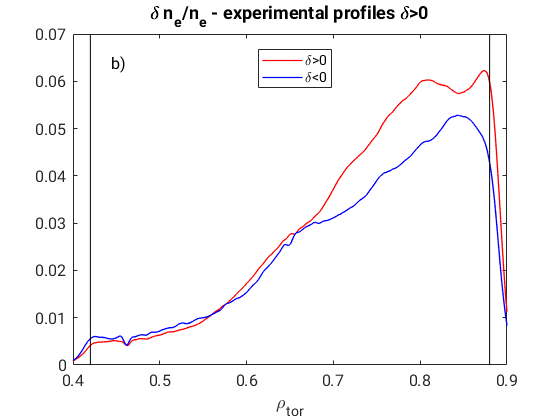}
    \caption{Same as Figure \ref{fig:global_Q_red}, but showing profiles of electron density fluctuations.}
  \label{fig:global_ne_red}
\end{figure}

\begin{figure}
    \centering
    \includegraphics[width=6cm]{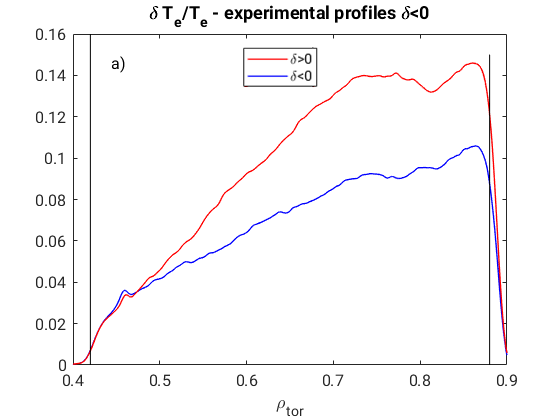}
    \includegraphics[width=6cm]{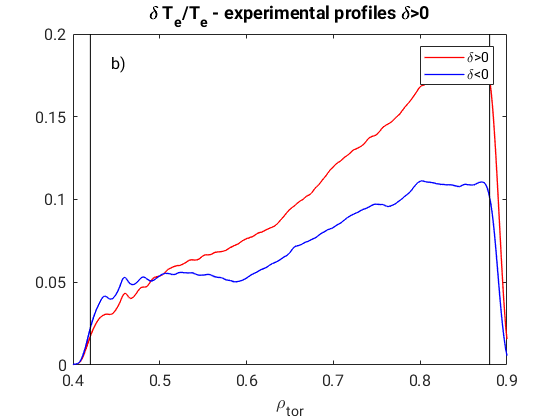}
    \caption{Same as Figure \ref{fig:global_ne_red}, but showing profiles of electron temperature fluctuation.}
  \label{fig:global_Te_red}
\end{figure}

The simulations presented in the previous section, while representing a successful example of comparison between global gyrokinetic simulations results and experimental plasmas, are not fully clarifying the basic impact of $\delta<0$ on transport. The main limitation is, as already mentioned,  the fact that they represent an experimental triangularity scan at constant power with different profiles for the two shapes, thus making it difficult to ascertain whether the effect of negative triangularity is universal or rather caused (or enhanced) by the profiles themselves. To at least partially overcome this limitation, we have carried out a series of runs mixing the experimental profiles and shapes. Thus, each TCV shape has been simulated considering the plasma profiles measured for both positive and negative $\delta$. To reduce the computational effort, this investigation has been carried out without considering impurities. Furthermore, since we are interested in the overall confinement behavior we plot only the total heat flow, the sum of the ion and electron channels. Results are summarized in Figure \ref{fig:global_Q_red} where we plot the overall transport level, while Figures \ref{fig:global_ne_red} and \ref{fig:global_Te_red} show the behavior of fluctuations.
One clearly sees that negative triangularity leads in all cases to reduced heat fluxes, regardless of the specific choice of profiles;  the absolute level of transport and fluctuations reduction is however dependent the profiles. Examining first the total transport level, Figure \ref{fig:global_Q_red}, we see that for both profiles the flux is about twice as large when $\delta>0$. In agreement with the results discussed in the previous section, the dominant transport channel is the electron one, responsible for approximately $75\%$ of the overall flux when considering positive triangularity experimental profiles and about $60\%$ when $\delta<0$. The magnitude of the fluxes clearly indicates that impurities play a major role in reducing the transport level. This is mostly evident comparing the results obtained with $\delta<0$ experimental profiles and Figure \ref{fig:global_transport}, which shows a reduction by almost a factor of two, mostly due to dilution of the main species and collisional stabilization of the TEM. A much smaller reduction is observed when the $\delta>0$ case  is considered. In this case including impurities does not alter significantly the total transport level which in fact increases in the core region ($\rho_{\rm tor} \simeq 0.5$) and  somewhat reduced in the near edge region ($\rho_{tor}\simeq0.8$). This is understood as a consequence of TCV being characterized by a mixed density/temperature driven TEM, making the transport level particularly sensitive also to variations in the background density gradient (in agreement with the local results of Ref.~\cite{Merlo_PPCF} and as also verified for these discharges by performing a few local simulations). 

Similarly to the heat flux, radial profiles of fluctuations also show a reduction when $\delta<0$. The exact magnitude of the reduction is case dependent: for density (Figure \ref{fig:global_ne_red}) one observes fluctuations being about 1.5 larger when $\delta>0$, over a radial domain that depends on the profiles (smaller for the experimental profiles of $\delta>0$). Similar reduction is observed for temperature fluctuations, Figure \ref{fig:global_Te_red}. In fact more than the actual fluctuation reduction, it is important to note that such a reduction happens almost uniformly over the region $\rho_{\rm tor}>0.6$ and does not follow the difference between the local values of triangularity.

It should nevertheless be pointed out that the results of this numerical experiment show again the limitations caused by the gradient driven approach and the turbulent transport being stiff. In a truly steady state one would expect the heat flow to be constant in any source free region. In our case where the ECH source is on axis, excluding radiative losses and the much smaller Ohmic heating, we would expect the heat flux to be constant as a function of radius throughout the entire plasma minor radius, similarly to what the experimental power balance shows. This is clearly not obtained for any of the simulations discussed in this Section, nor for the more ``realistic'' results presented in Sec.~\ref{sec:exp} including impurities. The fact that one obtains peaked flux profiles (see e.g. Figure \ref{fig:global_Q_red}, right panel) is in itself a sign that the specific profiles we have used are potentially not fully realistic or, equivalently, they can be obtained with a source that is not as localized as the experimental one. Nonetheless, this behaviour is systematic and if one was to experimentally obtain the profiles we have used as inputs, the transport and fluctuation reduction would be seen.

\section{Summary and conclusion}\label{sec:concl}
We have performed the first global nonlinear gyrokinetic simulations of TCV plasmas with positive and negative triangularity, making use of the GENE code. Simulation parameters were based on a limited L-mode plasma corresponding to an experimental triangularity scan at constant heating power. Nonlinear simulations are found to reproduce the experimental transport level for both shapes over a significant fraction of the plasma minor radius. A larger gradient is sustained in the $\delta<0 $ plasma, which leads to an improved electron energy confinement. Gyrokinetic simulation results are unable to recover the transport level in the deep core region ($\rho_{tor}<0.5$), where both global and local simulation show a much smaller turbulence level than experiments. Given that this region is very close to the inversion radius, this is understood as a consequence of difficulties in reconstructing the plasma profiles, as well as the known sensitivity of gradient driven simulations to input profiles. Simulation also reproduce an overall reduction of the electron density fluctuations, in agreement to what is experimentally observed. A reduction of $\delta T_e/T_e$ is also observed. It is however mostly due to the increased electron temperature in the $\delta<0$ plasma. In order to investigate the interplay between triangularity and plasma profiles, we have also carried out a series of simulations mixing the experimental profiles and shapes. Even though the results are affected by the transport being stiff, when considering the same profiles for both shapes, both fluxes and fluctuations are reduced when $\delta<0$. The reduction is global while its exact magnitude is profile dependent.\\
The importance of global effects has been addressed ny comparing global results to flux-tube simulations, and a very different behavior is observed. Surprisingly, for the cases we have considered very modest global effects are seen for the positive triangularity plasma, while a reduction in heat flux by approximately a factor of four is observed in the global run for the $\delta<0$ case, enough to reconcile simulations and experiments. Local simulations results are, as expected, very sensitive to variations of the electron temperature gradient, and while for $\delta>0$ it is possible to recover the measured transport level, this is not the case for $\delta<0$. Including global effects is thus necessary to reproduce the experimental transport level.\\
For future work, further comparisons between negative triangularity TCV plasmas and simulations are desirable. Such an effort, using dedicated synthetic diagnostics, can provide a unique validation opportunity of the gyrokinetic model in an unconventional regime. Furthermore flux-driven simulations should be attempted to alleviate the concerns of gradient-driven simulations regarding profile stiffness.  Aside from TCV, the recent observations reported from DIII-D clearly point out that the improvement associated to negative triangularity is n not the exclusive purview of small machines, but extends well beyond it, providing an exciting alternative for future reactors. Comparing and contrasting TCV and DIII-D simulation results will allow us to deepen our understanding of the confinement improvement as well as of global effects, and potentially help in optimizing future reactors. 

\section{Acknowledgements}
The authors acknowledge the Texas Advanced Computing Center (TACC) at The University of Texas at Austin for providing HPC resources that have contributed to the research results reported within this paper (Grant: DE-FG02-04ER54742). Part of the simulations have been performed on the Marconi supercomputer at CINECA. This work was supported in part by the Swiss National Science Foundation.

\section*{References}
\bibliographystyle{iopart-num}
\bibliography{bibliography}

\end{document}